\documentclass[twocolumn]{article}

\usepackage[applemac]{inputenc}
\usepackage[T1]{fontenc} 
\usepackage[english]{babel}
\usepackage{amsmath,amssymb,amsthm}
\usepackage{natbib}
\usepackage{graphicx,tikz}
\usepackage{array,multirow}
\usepackage{caption, subcaption}
\usepackage{floatrow}
\usepackage{hyperref}

\title{Reliability Gaps Between Groups in COMPAS Dataset}

\author{Tim Räz\footnote{University of Bern, Institute of Philosophy, L\"anggassstrasse 49a,  3012 Bern, Switzerland. e-mail: tim.raez@posteo.de}}

\begin{document}

\maketitle

\begin{abstract}
This paper investigates the inter-rater reliability of risk assessment instruments (RAIs). The main question is whether different, socially salient groups are affected differently by a lack of inter-rater reliability of RAIs, that is, whether mistakes with respect to different groups affects them differently. The question is investigated with a simulation study of the COMPAS dataset. A controlled degree of noise is injected into the input data of a predictive model; the noise can be interpreted as a synthetic rater that makes mistakes. The main finding is that there are systematic differences in output reliability between groups in the COMPAS dataset. The sign of the difference depends on the kind of inter-rater statistic that is used (Cohen's Kappa, Byrt's PABAK, ICC), and in particular whether or not a correction of predictions prevalences of the groups is used.
\end{abstract}

\section{Introduction}

We all make mistakes. However, mistakes have consequences, and the consequences are not the same for everybody. If we grade the exam of an excellent student incorrectly by counting a correct answer as wrong, their final mark may understate their achievement. If we grade the exam of a bad student incorrectly by counting a correct answer as wrong, they may have to repeat the grade, or not graduate at all. The present paper investigates the consequences of errors in the context of risk assessment in criminal justice, specifically, the inter-rater reliability of risk assessment instruments (RAIs). Risk assessment instruments take features of individuals as inputs, and output a risk score or risk prediction. The outputs may be used to, say, decide whether offenders are released on parole. The inputs of RAIs are often compiled by humans, and may thus contain mistakes. Inter-rater reliability measures whether different raters arrive at different ratings of the same individuals, and how these different ratings affect risk predictions.

The main question is whether there are systematic differences in inter-rater reliability between different, socially salient groups, that is, whether making mistakes with respect to groups affects them differently. It is well-known that there are historically entrenched socio-economic inequalities between different races. The consequences of these inequalities in risk assessment have been widely discussed, starting with the investigation of COMPAS by ProPublica \citep{angwi2016}. This discussion has primarily focused on predictive validity. Little attention has been paid to reliability. The present paper highlights the importance of (inter-rater) reliability, which can be interpreted as the absence of noise between raters. Reliability, like validity, is an important property of predictive models, but much less explored than the latter notion; on noise see \citet{kahne2022}.

Simulations based on the so-called COMPAS dataset are performed to investigate group-specific reliability differences. In recidivism risk assessment, the study of inter-rater reliability is usually based on empirical data. At least two human raters are tasked with rating a set of individuals. The resulting ratings are used to calculate well-established inter-rater statistics for both ratings and risk predictions. However, there is litte actual inter-rater data available, and existing datasets are often small, which limits their usefulness, in particular if one wants to study the reliability of subsets of a sample (groups). It is proposed here that in view of scarce empirical data, simulation experiments are useful. In the simulations, a controlled degree of noise is injected into the input data of a predictive model; the noise can be interpreted as a synthetic rater that makes a certain amount of mistakes. The reliability of the resulting predictions is then measured. This helps to gauge how different groups are affected by the same amount of noise.

The main finding of the paper is that there are systematic reliability differences between group 0 (non-White people) and group 1 (White people) in the COMPAS dataset. The sign of the difference depends on the kind of inter-rater statistic that is used. Common inter-rater statistics, an intraclass correlation coefficient, as well as Cohen's Kappa, find that group 1 has mostly lower levels of reliability. However, a different inter-rater statistic, Byrt's PABAK (also known as Bennett's S), finds that group 0 has mostly lower levels of reliability. To interpret these findings, a well-known fact about Cohen's Kappa is crucial, viz. that it is a prevalence-dependent measure of reliability. If a group has a lower prevalence (base rate) of predicted risk, the reliability of that group will be underestimated by Cohen's Kappa in comparison to the other group. The higher prevalence of predicted risk for group 0 is a well-known consequence of historical inequalities. Thus, it is found that if we measure reliability for these groups \emph{without correcting for different prevalences}, then we find that group 1 has predictions with lower reliability than group 0, for the same amount of noise. The finding is reversed if we use a prevalence correction, in which case group 0 has predictions with lower reliability.

\section{Background}
\label{sec:background}

\subsection{Inter-rater Reliability (IRR)}

Reliability is a well-known desideratum of recidivism risk assessment  instruments in criminal justice \citep{desma2018}. In the present paper, the focus is on inter-rater reliability. In empirical studies, inter-rater data is collected by choosing a sample of individuals of adequate size and composition. Adequately trained raters are then tasked with compiling ratings for these individuals. These ratings, and risk predictions based on these ratings, are then assessed with respect to reliability. In the present paper, IRR with respect to two raters, $r$ and $s$, will be considered. Both raters rate a set of individuals $\{1, ..., n\}$ with respect to $j$ items. The rating of individual $i \in \{1, ..., n\}$ by rater $k$ on the $j$ items can be written as $x_k(i) = (x_{k1}(i), x_{k2}(i), ..., x_{kj}(i))$. Thus, a full set of ratings for two raters has the form of table \ref{irr_data}.

\begin{table}[!h]
\begin{center}
\begin{tabular}{|c|c|c|}
\hline
ind. & rater r & rater s \\
\hline
1 & $(x_{r1}(1), ..., x_{rj}(1))$ & $(x_{s1}(1), ..., x_{sj}(1))$ \\
2 & $(x_{r1}(2), ..., x_{rj}(2))$ & $(x_{s1}(2), ..., x_{sj}(2))$ \\
... & ... & ... \\
n & $(x_{r1}(n), ..., x_{rj}(n))$ & $(x_{s1}(n), ..., x_{sj}(n))$ \\
\hline
\end{tabular}
\end{center}
\caption{Inter-rater reliability data: raters $r$ and $s$ rate items $1, ..., j$ for individuals $1, ..., n$.}
\label{irr_data}
\end{table}

On the basis of table \ref{irr_data}, one can compute the IRR with respect to individual items. In the present paper, data as in table \ref{irr_data} will be simluated. Starting with one set of ratings, the column `rater $r$', the column `rater $s$' will be generated by adding noise to `rater $r$'. Only the IRR of the \emph{predictions} will be computed. One obtains predictions $\hat{y} = f(x_k(i))$ from the ratings in table \ref{irr_data} using a RAI, represented by $f:X \rightarrow \hat{Y}$. Then one computes the reliability of the predictions. Data suitable to compute the reliability of predictions take the form of table \ref{irr_predictions}.

\begin{table}[!h]
\begin{center}
\begin{tabular}{|c|c|c|}
\hline
ind. & rater r & rater s \\
\hline
1 & $f(x_r(1))$ & $f(x_s(1))$ \\
2 & $f(x_r(2))$ & $f(x_s(2))$ \\
... & ... & ... \\
n & $f(x_r(n))$ & $f(x_s(n))$ \\ 
\hline
\end{tabular}
\end{center}
\caption{Predictions based on ratings $r$ and $s$ of individuals $1, ..., n$.}
\label{irr_predictions}
\end{table}

The IRR statistic to be used depends on the kind of prediction. In the case of binary predictions, Cohen's Kappa, or a variant thereof, is appropriate. In the case of continuous predictions, a version of an intraclass correlation coefficient (ICC) is appropriate. These statistics will now be introduced.

\subsection{Cohen's Kappa (CK)} 

For binary predictions ($0$ and $1$), inter-rater reliability statistics can be computed from a confusion matrix. The confusion matrix is a summary of table \ref{irr_predictions}. Let $a$ be the number of rows with the pair $(1, 1)$, $b$ the number of rows with $(0, 1)$, $c$ the number of rows with $(1, 0)$, and $d$ the number of rows with $(0,0)$. The confusion matrix is:

\begin{figure}[h!]
\centering
\begin{tabular}{l|c|c|c|c}
\multicolumn{2}{c}{}&\multicolumn{2}{c}{rater r}&\\
\cline{3-4}
\multicolumn{2}{c|}{}&1&0&\multicolumn{1}{c}{total}\\
\cline{2-4}
\multirow{2}{*}{rater s}& 1 & $a$ & $b$ & $s_1$\\
\cline{2-4}
& 0 & $c$ & $d$ & $s_0$\\
\cline{2-4}
\multicolumn{1}{c}{} & \multicolumn{1}{c}{total} & \multicolumn{1}{c}{$r_1$} & \multicolumn{1}{c}{$r_0$} & \multicolumn{1}{c}{$n$}\\
\end{tabular}
\end{figure}

A first, simple reliability statistic is \emph{observed agreement}, $p_o = (a+d)/n$. Observed agreement is the proportion of cases in which the two raters agree. Observed agreement is easy to measure and interpret, but has the disadvantage that it does not correct for the possibility of chance agreement, that is, agreement that we would expect even if the two raters make random predictions. A statistic that solves this problem is Cohen's Kappa, (CK), proposed by \citet{cohen1960}. CK measures agreement between two (or more) raters, where we correct observed agreement with (expected) agreement by chance. \emph{Agreement by chance} is the proportion of cases in which we expect the two raters to agree at random, estimated from the raters' base rates. To obtain agreement by chance, we multiply the proportions by which the two raters assign decision subject to either $1$ or $0$ (rater prevalences), finding agreement by chance as $p_c = (r_1\cdot s_1 + r_0 \cdot s_0)/n^2$. Cohen's Kappa now takes the form

\begin{equation}
CK = \frac{p_o - p_c}{1 - p_c},
\end{equation}

where $p_o$ is observed agreement. CK lies in $[-1, 1]$. A value of $1$ means perfect agreement, $0$ means agreement is purely by chance, and $-1$ means perfect disagreement. Cohen's Kappa is one of the most frequently reported IRR statistics. However, two well-known problems are associated with it, cf. \citet{hallg2012}. The first problem is the \textit{prevalence problem}. If the marginal distribution of predictions is unbalanced, i.e., much lower for, say, $1$ than for $0$, then CK may be too low, i.e., it may underestimate reliability. The second problem is the \textit{bias problem}. If the marginal distribution of the two raters is substantially different, then CK may be too high, i.e., it may overestimate reliability. In these two situations, CK may mis-represent inter-rater reliability.

\subsection{Prevalence-Adjusted Bias-Adjusted Kappa (PABAK)}

Different remedies for the two problems have been recommended in the literature \citep{hallg2012}. Here the approach proposed by \citet{byrt1993} is adopted. The idea is to compute a corrected version of Kappa called ``Prevalence-Adjusted Bias-Adjusted Kappa'' (PABAK), which takes both prevalence and bias into account. PABAK corrects for unequal prevalences by computing a Kappa statistic in which both positive and negative agreement, $a$ and $d$, are replaced by their average $(a+d)/2$. At the same time, both disagreement statistics, $b$ and $c$, are replaced by their average $(b+c)/2$ to correct for rater bias. The resulting statistic, PABAK\footnote{Note that PABAK is identical with Bennett's S, but with a different motivation \citep{byrt1993}.}, is defined as:

\begin{equation}
PABAK = \frac{2(a+d)}{n} - 1 = 2p_o - 1.
\end{equation}

PABAK is a linear rescaling of observed agreement ($p_o$) and lies in $[-1, 1]$. PABAK corrects for bias and prevalence at the same time, and as a consequence, it is not clear which correction is responsible for differences between CK and PABAK. To solve this problem, \citet{byrt1993} propose to monitor both bias and prevalence, measured as the ``Bias Index'' (BI) and the ``Prevalence Index'' (PI). The Bias Index is defined as the difference in proportions of disagreements. In terms of the confusion matrix, $BI = (b-c)/n$. The Prevalence Index is defined as the difference in proportions of positive agreements and negative agreements. In terms of the confusion matrix, we define: $PI = (a-d)/n$.\footnote{Bias and prevalence as defined here are not necessarily standard uses of these terms. A difference in prevalences between groups can also be interpreted as a form of (historical) bias: Different, socially relevant groups have different prevalences of, e.g., recidivism due to historically entrenched inequalities, cf. \citep{raez2021}. The technical notion of bias defined here is a kind of rater bias, because it measures an inequality in rater base rates.} Byrt et al. establish the following relation between Cohen's Kappa and PABAK:

\begin{equation}
CK = \frac{PABAK - PI^2 + BI^2}{1 - PI^2 + BI^2} \label{pabak_k}
\end{equation}

This relation between CK and PABAK is relevant because, first, it supports the above observation that PABAK corrects for bias and prevalence, where imbalanced prevalences leads to an underestimation of inter-rater agreement in CK (negative sign of $PI$ term), while bias leads to an overestimation of inter-rater agreement in CK (positive sign of $BI$ term). Second, it allows us to understand the source of disagreements between CK and PABAK by monitoring $PI$ and $BI$.

\subsection{Intraclass Correlation Coefficient (ICC)}

To measure inter-rater reliability with respect to continuous predictions such as scores and probabilities, other inter-rater reliability statistics are needed. The most important family of such statistics are so-called ``Intraclass Correlation Coefficients'' (ICCs); see \citet{lilje2019} for a useful overview and discussion. ICCs take values in $[0, 1]$. ICCs can be described as ratios of variances, (variance of interest)/(total variance) \citep{lilje2019}. In the context of inter-rater reliability, the variance of interest includes differences between the ``true'' ratings of different individuals, which we expect, because different individuals have different underlying risk profiles. The total variance includes the variance of interest, but also differences between raters (rater bias), and differences between different ratings of the same individuals. If the total variance is large in comparison to the variance of interest, ICC will be close to $0$. If the total variance is similar to the variance of interest, ICC will be close to $1$. 

An appropriate version of ICC has to be chosen based on the experimental situation. Here we draw on the flowchart in \citet[Fig. 1]{koo2016} to find the appropriate ICC. First, one has to determine whether or not the same raters rate all subjects, or if raters are assigned randomly to subjects. In our case, the same two (synthetic) raters $r$ and $s$ rate all subjects, cf. table \ref{irr_predictions}. This means that a so-called two-way model is appropriate. Second, these (synthetic) raters have specific and unique properties (known noise levels), which means that a so-called mixed effects model is appropriate. Third, we use two single (synthetic) raters $r$ and $s$, rather than means of different raters, such that a so-called single rater model is appropriate. Fourth, and finally, we are interested in so-called absolute agreement rather than consistency. Absolute agreement measures whether different raters assign the same rating to the same subjects. Consistency measures whether different raters assign the same ratings to the same subjects with an additive correction. This means that for consistency, the raters have to agree on the order of ratings, but their overall rating level can differ.\footnote{For example, if $r$ and $s$ assign different ratings to subjects $i, i'$, i.e., $x_r(i) \neq x_s(i)$ and $x_r(i') \neq x_s(i')$, then absolute agreement is violated. However, consistency can still be satisfied if there is additive rater bias, i.e., if there is a constant $c$ such that $x_r(i) = x_s(i) +c$ and $x_r(i') = x_s(i') +c$, cf. \citep{koo2016}, i.e., $r$ and $s$ agree on the relative order of $i$ and $i'$.} For present purposes, consistency is not appropriate because scores are usually transformed into risk levels or (binary) decisions using thresholds, and for this transformation, the absolute value of the score is used. Two classic papers \citep{shrou1979,mcgra1996} propose different ways of reporting ICCs. Following the convention of \citet{mcgra1996}, $ICC(A,1)$, a two-way, single rater ICC for absolute agreement is measured.\footnote{$ICC(A, 1)$ has different interpretations for the mixed effects and the random effects case, but the computation is the same, as \citet{lilje2019} point out.}

\section{Related Work}
\label{sec:relatedwork}

Reliability is a very general desideratum of (predictive) modeling. Statistical methods for measuring reliability have been developed and applied in medicine and biology as well as social sciences and psychology \citep{baner1999, lilje2019}. The importance of reliability also applies to recidivism risk assessment instruments, where it is often mentioned as one of two important properties of such instruments, to be evaluated alongside predictive validity, cf. \citep{desma2018}. There is a lack of knowledge regarding predictive validity and reliability of risk assessment instruments (Ibid.), and this is particularly true for (inter-rater) reliability. In a survey of risk assessment instruments in the US, \citet{desma2018} included 53 publications and evaluated 19 risk assessment instruments. In the results section, they write: ``Perhaps one our most striking findings, only two of the 53 studies reported on the inter-rater reliability of the risk assessments'' (Ibid., p. 20). Desmarais et al. continue that there is a ``critical need for data on the inter-rater reliability'' of recidivism risk assessment instruments in the US. \citet{baird2009} provides an even more scathing assessment of reliability research: ``Nearly all of the literature on popular risk models refers to their demonstrated validity and reliability. In actuality, there is little information available that supports model reliability, and much of what is available either addresses the wrong issue (internal consistency) or provides inadequate tests of inter-rater reliability''. 

Below, data on COMPAS, one of the most high-profile risk assessment instruments, will be used. The evaluation of COMPAS by ProPublica \citep{angwi2016} constitutes one of the starting points of the debate on fairness in machine learning \citep{baroc2019}. The predictive validity of COMPAS has been empirically investigated in different studies, cf. \citep{brenn2018}. In contrast, only one (test-retest) reliability study of COMPAS has ever been conducted \citep{farab2010}.\footnote{The author tried to obtain the data used in \citep{farab2010}. Please contact the author if you have access to the data, or know how to gain access.} \citet{brenn2018} write that two studies of IRR for COMPAS are planned. However, apparently, these two studies studies were never completed and/or published. Thus, the inter-rater reliability of one of the most high-profile risk assessment instruments in use today has never been empirically investigated. The reason for the lack of studies may be that collecting suitable data is expensive, because at least two raters instead of one have to compile ratings, study case files, conduct interviews, and so on. In the case of  COMPAS, it is reported that rating a single individual may take between 10 to 60 minutes \citep[p. 7]{desma2018}, which shows how time-consuming measuring IRR can be. Presumably, institutions such as correctional facilities that use RAIs do not have the resources to conduct large IRR studies.

Empirical studies of validity and reliability of risk assessment instruments with respect to socially relevant groups have been conducted. However, the author is not aware of empirical studies that have found systematic inter-rater reliability differences between socially relevant groups. \citet{holtf2007} provide a review of gender-specific research of LSI-R, an important RAI. They find that few studies examine validity or reliability separately for females. While one study reported high IRR (percent agreement) for a sample of young women, there appears to be no study considering gender differences in reliability. \citet{stewa2011} investigate validity and reliability of different RAIs, including LSI-R, for female-only samples. \citet{lowde2019} examine reliability and validity of START and LSI-R, two risk assessment instruments. While they investigate (predictive) validity by race, and measure ICC and Cohen's Kappa for total scores and risk bins, they do not report reliability results by race. \citet{jimen2018} investigate the LS/CMI and examine whether total scores and raters yielded differences between majority and minority groups. They find that there are group-specific differences of ratings (higher scores for minorities on seven of eight criminogenic factors), but that there is no evidence that this difference is due to racial bias in the administration of the instrument (rating). Jimenez et al. did not investigate rater bias via inter-rater reliability statistics. \citet{lowde2019a} investigate racial bias with respect to LSI-R assessments. However, inter-rater reliability was not investigated due to a lack of data. Rater cultural bias was explored in \citet{venne2021}. Venner et al. found no evidence of rater cultural bias for YLS/CMI-SRV.

The use of machine learning methods for risk assessment in criminal justice has been the focus of renewed interest at least since \citet{angwi2016} and the discussion of COMPAS.\footnote{Note that COMPAS is based on a so-called logistic regression model with $l_1$ regularization, i.e., well-known methods from machine learning and statistics, cf. \citealt{brenn2018}. A very similar model will be used in the simulation study below.} Since the publication of Angwin et al., many issues of ML methods in application to risk assessment have been discussed; the main issues have been predictive validity (in particular the possibility of enhancing predictive performance with ML methods), fairness, transparency, and various tradeoffs between these desiderata. However, the issue of (inter-rater) reliability is conspicuously absent from the debate. For example, \citet{kiger2022}, a paper on ML methodology for risk assessment, focuses on predictive performance; Kigerl et al. do not mention (inter-rater) reliability. Similarly, \citet{trava2022}, a recent review of ML methodology for recidivism risk assessment, provides comparisons of predictive validity of different ML methods and datasets, but does not discuss (inter-rater) reliability as a desideratum of ML-enhanced risk assessment.

\section{Methods}
\label{sec:methods}

The code for all experiments, results and figures can be found on the following github repository: \href{https://github.com/timraez/ReliabilityGapsCOMPAS}{timraez/ReliabilityGapsCOMPAS}.

\subsection{Outline of Experiments}
\label{sec:idea}

The key idea of the empirical part is to simulate inter-rater data. Any dataset suitable to test the predictive validity of a RAI with information about group membership can serve as a basis. Access to a RAI or a predictive model is necessary. Access to the labels of test data and inter-rater data are not necessary. In more detail, we need a dataset $X_{orig}$ of single ratings of $n$ individuals, which has the form of column `rater r' of table \ref{irr_data}. We assume that each individual belongs to exactly one of two groups ($a$ or $b$). The predictive model (RAI) $f:X \rightarrow \hat{Y}$ should provide predictions for entries of $X_{orig}$. The basic simulation experiment has the following steps:

\begin{enumerate}

\item Perturb the data $X_{orig}$:
\begin{equation*}
X_{pert} = P(X_{orig}) 
\end{equation*}

The perturbed data $X_{pert}$ corresponds to the ratings of a second rater, i.e., the column `rater s' of table \ref{irr_data}. The kind of perturbation will depend on the form of items (binary or numerical).

\item Apply the predictive model (RAI) $f$ to original and perturbed data (possibly after preprocessing) to obtain predictions:
\begin{eqnarray*}
\hat{Y}_{orig} &=& f(X_{orig}), \\
\hat{Y}_{pert} &=&f(X_{pert}).
\end{eqnarray*}
Predictions can be binary or risk scores. The resulting data corresponds to table \ref{irr_predictions}.

\item Split the predictions according to group membership ($a$ or $b$) of individuals to obtain group-specific predictions:
\begin{eqnarray*}
\hat{Y}^a_{orig} \cup \hat{Y}^b_{orig} &=& \hat{Y}_{orig},\\
\hat{Y}^a_{pert} \cup \hat{Y}^b_{pert} &=& \hat{Y}_{pert}.
\end{eqnarray*}

\item Pair original and perturbed predictions group-wise, with predictions for each individual on the original and the perturbed data, and compute the inter-rater reliability statistics $IRR$ separately for the two groups $a, b$
\begin{eqnarray*}
IRR^a &=& IRR \big( \hat{Y}^a_{orig}, \hat{Y}^a_{pert}\big)\\
IRR^b &=& IRR \big( \hat{Y}^b_{orig}, \hat{Y}^b_{pert}\big)
\end{eqnarray*}
The reliability statistic $IRR$ is chosen based on the type of prediction: Cohen's Kappa (or variants thereof) for binary predictions; intraclass correlation coefficients for risk scores.

\item Examine the difference between reliability statistics for the two groups, i.e., the difference between $IRR^a$ and $IRR^b$.

\end{enumerate}

\subsection{Data}

Experiments were performed on the COMPAS dataset, compiled by ProPublica \citep{angwi2016} to investigate issues of group fairness with COMPAS. The version (and setup) of \citet{hertw2022} was used, which, in turn, is based on \citet{fried2019}. The COMPAS dataset contains $n=6167$ entries of defendants form Broward County, FL, USA, for which COMPAS risk scores were available. Ground truth is part of the dataset; the label $Y=1$ encodes actual rearrest within 2 years, while $Y=0$ encodes no actual arrest within two years. The sensitive attribute is race: $2100$ defendants are assigned to the category ``White'' (privileged group), encoded as $1$, while $4067$ defendants were assigned to the category ``non-White'' (non-privileged group), encoded as $0$. The dataset contains 14 features, including $Y$. Of the 13 input features, 8 are categorial (``misdemeanor charge'', ``felony charge'', ``sex'', ``race'', ``sex-race'', 3 age categories), and 5 are numerical (``age'', ``juvenile felony charges count'', ``juvenile misdemeanor charges count', ``juvenile other charges counts'', ``priors count''). To keep perturbations simple, the 3 one-hot encoded age categories were used in training and for predictions, but they were not perturbed. ``race'' and ``sex-race'' were also not perturbed in order to not confound group membership. Note that the COMPAS dataset contains features that can be used to predict recidivism, but it does not contain input data for COMPAS (the COMPAS questionnaire contains over 130 items). Eventually, experiments should be performed on actual risk assessment data, and with actual RAIs. The experiments reported here should be interpreted as proof-of-concept, demonstrating the in-principle feasibility of the method. 

\subsection{Data Split and Model}

The dataset was split into 5 folds, as for 5-fold cross-validation. A standard logistic regression model (with automatic $l_2$ regularization and L-BFGS solver) was fit to each of the 5 train splits, obtaining 5 models in total. This model serves as a stand-in for a RAI. Before fitting the model, a standard data normalization (scaling) was performed on the numerical features, as required for regularization. The scaling was applied to perturbed and unperturbed test data before obtaining predictions. There was no further optimization or parameter tuning of the 5 models. All subsequent experiments were performed with the 5 models on the 5 test splits. Reported results were obtained as averages of the 5 splits to increase stability.

\subsection{Perturbations}

For each experiment, perturbed copies of the test sets were obtained by applying the following procedure to each of 5 test sets from the 5 folds. The original test set $X_{orig}$ was copied, and a perturbation (noise) $P$ was added to the copy to obtain a perturbed version $X_{pert} = P(X_{orig})$. Experiments were performed with noise of different strengths and on different subsets of variables. For categorial (binary) variables, values were flipped ($0 \mapsto 1, 1 \mapsto 0$) at random with a given probability (noise level) $p$. This was achieved by adding a random vector with a proportion $p$ of $1$s to the binary data, and then taking the result mod 2. For numerical variables, normal noise with $\mu = 0$ and variance $\sigma^2$ was added with a given probability (noise level) $p$. This was achieved by generating a random vector with draws from the normal distribution $\mathcal{N}(0, \sigma^2)$, rounded to the closest integer. The resulting vector was multiplied with a vector of $0$s and $1$s, with a proportion $p$ of $1$s, thus obtaining a vector with a proportion $p$ of entries drawn from $\mathcal{N}(0, \sigma^2)$ and $0$ elsewhere, which was added to the data. In the experiments described below, noise was added with the following parameter ranges: $p \in [0, 0.3]$, with increments of $0.01$, and $\sigma^2  \in \{1, 5, 10\}$. The parameter $p$ controls for the level of noise, while $\sigma^2$ determines the size of errors for numerical variables. For example, the combination of $p = 0.2, \sigma^2 = 5$ means that noise was added to a fraction of $.2$ of inputs in question, and in the case of numerical variables, this noise has distribution $\mathcal{N}(0, 5)$. Each such combination of values corresponds to a basic experiment as described in Sec. \ref{sec:idea}.

This noise scheme was chosen to create a level of noise (proportion of perturbed values) that could be varied continuously for categorical and numerical variables at the same time.\footnote{Note that the level of noise for numerical variables is an upper bound, because draws from, e.g., $\mathcal{N}(0, 1)$ will be $0$ with a probability of .38.} Of course, this noise may be unrealistic. In a first set of experiments, no scaling was used to prevent negative values for numerical features. Depending on the RAI in question, negative values for, e.g., the count of priors, or values for age below 18, are excluded \emph{a priori}. Further experiments with such minima for numerical variables were performed. 

If noise is applied to one or several features without taking group membership into account, the noise is independent of group membership (in expectation). However, a certain random fluctuation will be introduced by not controlling for how the noise is distributed over the two groups. Therefore, the addition of noise was performed separately for each group, using identical noise levels. It was found that controlling for noise distribution over groups yielded more stable results, but did not have a qualitative impact, as would be expected for large enough samples. Note that adding noise in this way can be interpreted as guaranteeing that the second rater makes mistakes without group bias.

\subsection{Interpretation of IRR Values, Statistical Significance, and Qualitative Differences}

One of the difficulties of measuring IRR is the interpretation of the (absolute) values of IRR statistics. The range of inter-rater statistics do not have precise interpretations (except for extrema, see above). It has become common practice to report certain levels of these statistics as excellent, good, acceptable, etc. . However, saying that, e.g., ICC reliability in the interval $[0.9, 1]$ is ``excellent'', as suggested in \citet{koo2016}, is purely conventional, arbitrary, and lacks justification. For this reason, such qualitative evaluations will not be reported here. Also, the focus of the present paper is on systematic differences of values of these statistics for different groups, which means that absolute values are less significant.

What does it mean that the difference between the values of two groups is systematic? It can mean that the difference between the values of two groups is statistically significant. For some of the statistics considered here (Cohen's Kappa, ICC), test of statistical significance and confidence intervals are available and can be easily computed. In the case of simulation experiments, we can examine group differences between IRR statistics over a range of noise levels, as opposed to empirical studies, which usually only report point measurements (one value per IRR statistic and group). The different noise levels can be interpreted as a range of different synthetic raters. This makes it possible to determine qualitative differences. If the values of one group are always or mostly lower than the values of the other group for the entire set of synthetic raters (all levels of noise), this indicates that there are qualitative differences in reliability, even if the absolute differences between groups is small. Of course, the absolute difference has to be taken into consideration. If it is very small, a systematic difference could be negligible in practice.

\section{Results}
\label{sec:results}

\begin{figure*}[ht] 
     \centering
     \begin{subfigure}{0.3\textwidth}
         \centering
         \includegraphics[width=\textwidth]{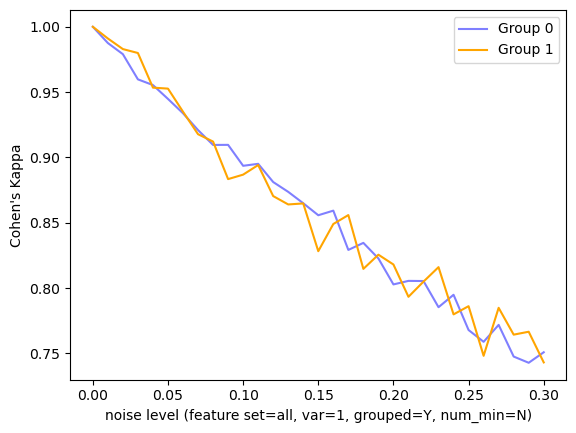}
         \caption{Cohen's Kappa}
         \label{fig:CK_s1}
     \end{subfigure}
     \hfill
     \begin{subfigure}{0.3\textwidth}
         \centering
         \includegraphics[width=\textwidth]{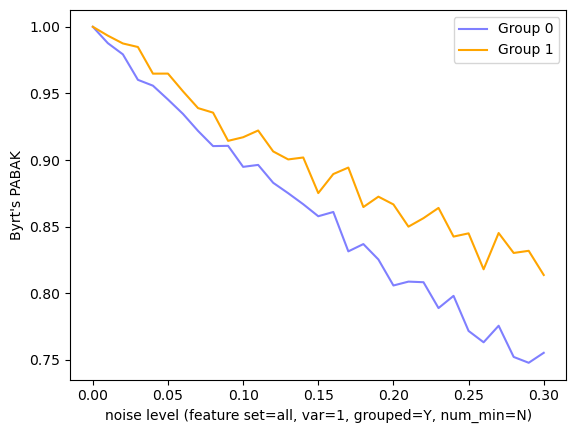}
         \caption{Byrt's PABAK}
         \label{fig:PABAK_s1}
     \end{subfigure}
          \hfill
     \begin{subfigure}{0.3\textwidth}
         \centering
         \includegraphics[width=\textwidth]{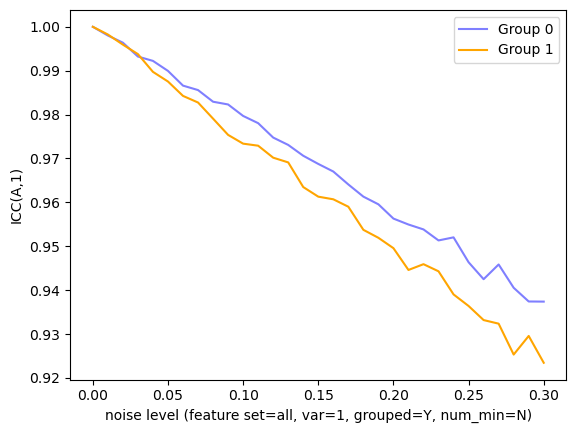}
         \caption{ICC(A,1)}
         \label{fig:ICCA_s1}
     \end{subfigure}
        \caption{Experiment 1,  $\sigma^2 = 1$.}
\end{figure*}

The goal of the experiments is to examine the overall effect of noise on group differences. The level of noise $p$ was increased from $0$ to $0.3$ in increments of $0.01$. Different kinds of noise were used for categorial and numerical features, as explained in the methods section. For the numerical features, three settings for the error size, measured as the variance of centered normal noise ($\mu = 0, \sigma^2 \in \{1, 5, 10\}$), were examined. The same level of noise was injected separately for the two groups. All features were perturbed at once with the same level of noise. Perturbed categorial features: ``misdemeanor charge'', ``felony charge'', ``sex''. Not perturbed: ``race'', ``sex-race'', and age bins. Perturbed numerical features: ``age'', ``juvenile felony charges count'', ``juvenile misdemeanor charges count', ``juvenile other charges counts'', ``priors count''. Additional experiments are summarized in Sec. \ref{sec:further_experiments}; the corresponding figures can be found in the appendix.

\subsection{Experiment 1, Perturbation of All Features, $\sigma^2  = 1$}
\label{exp_all_s1}

In this experiment, the error size for the numerical variables is small -- of the proportion $p$ of values that are perturbed, over $68\%$ are either $0$ or $1$. We can see in Fig. \ref{fig:CK_s1} that Kappa decreases for both groups as the noise level rises, as expected. There is no systematic reliability difference between the two groups. PABAK reliability decreases as the level of noise rises (Fig. \ref{fig:PABAK_s1}). There is a systematic difference between groups: Group 0 has lower reliability levels than group 1 for the same noise. The injection of the same level of noise affects different groups differently, such that group 0 achieves lower reliability. This is one of the main findings of the present paper. ICC(A, 1), in Fig. \ref{fig:ICCA_s1}, shows a different picture than both CK (Fig. \ref{fig:CK_s1}) and PABAK (Fig. \ref{fig:PABAK_s1}). The ICC reliability of group 1 is lower for same level of noise than the ICC of group 0.

\begin{figure*}[!h] 
     \centering
     \begin{subfigure}{0.4\textwidth}
         \centering
         \includegraphics[width=\textwidth]{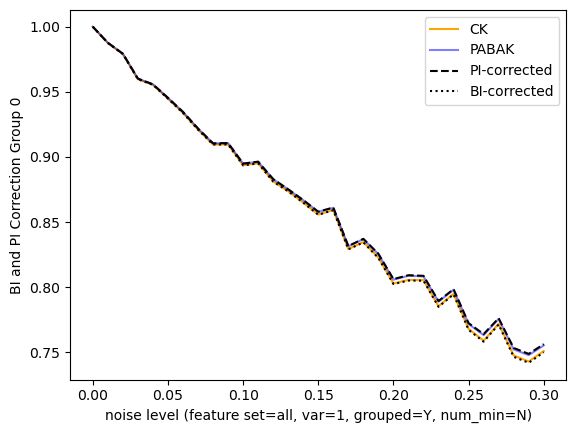}
         \caption{Group 0}
         \label{fig:corr_g0_s1}
     \end{subfigure}
     \hfill
     \begin{subfigure}{0.4\textwidth}
         \centering
         \includegraphics[width=\textwidth]{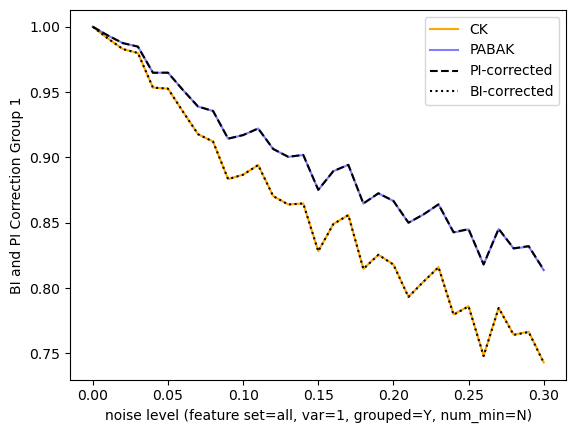}
         \caption{Group 1}
         \label{fig:corr_g1_s1}
     \end{subfigure}
        \caption{PI and BI correction, Experiment 1, $\sigma^2 = 1$.}
\end{figure*}

Now let us try to understand the difference between the results according to CK (Fig. \ref{fig:CK_s1}) and PABAK (Fig. \ref{fig:PABAK_s1}). We know that the difference between CK and PABAK is due to the combination of bias and prevalence correction, because PABAK and CK are related via equation (\ref{pabak_k}). Prevalence and bias correction have a marked impact on reliability. While there is no systematic group difference according to CK, PABAK finds lower levels of reliability for group 0. But is the difference due to the bias correction, to the prevalence correction, or both? We can answer this question by decomposing the difference between PABAK and Cohen's Kappa using equation (\ref{pabak_k}). Specifically, we can examine, for each group separately, what happens if we only use a bias correction or a prevalence correction by setting either BI or PI in equation (\ref{pabak_k}) to $0$, and comparing the result to both CK and PABAK. The result for group $0$ is in figure \ref{fig:corr_g0_s1}. We see that there is not a big difference between CK and PABAK. We also observe that if we only use the BI correction in equation (\ref{pabak_k}), that is, if we let PI = 0, the result largely coincides with CK. If we only use the PI correction by letting BI = 0, the result largely coincides with PABAK. This means that the prevalence correction is responsible for the (small) difference between CK and PABAK of this group. Now consider the result for group $1$ in figure \ref{fig:corr_g1_s1}. For this group, we observe that there is a bigger difference between CK and PABAK. This difference is responsible for the difference between figures \ref{fig:CK_s1} and \ref{fig:PABAK_s1}. We also observe that if we only use bias correction (PI = 0)  in equation (\ref{pabak_k}), the result again coincides with CK, whereas if we use prevalence correction (BI = 0), the result coincides with PABAK. This means that the difference between CK and PABAK is again largely due to the prevalence correction. CK underestimates the reliability for group 1 because this group has a lower prevalence than group $0$. The correction can be observed in Fig. \ref{fig:corr_g1_s1}.

A third important finding is that the results of ICC and PABAK contradict each other. How can this be explained? For one, ICC is measured with respect to scores, which means that we need not expect the same results as for PABAK.\footnote{The fact that ICC is calculated from scores may also explain why the ICC graph is smoother than graphs of binary predictions.} A plausible explanation of the difference is that PABAK is prevalence-corrected, while ICC does not take prevalences into account. Of course, we cannot directly measure (binary) prevalences for probabilistic predictions, but we can measure and compare average predicted risk scores for the two groups (not reported here). These group-specific average scores show a similar picture as binary prevalences: Group 0 has a higher average score than group 1, and this difference persists over the range of $p$, although the difference gets somewhat smaller as $p$ increases. Thus, the lower ICC reliability level of group $1$ may be due to the lower average score of this group. It should be stressed that this explanation is only a conjecture. It is unclear how prevalence affects ICC on a theoretical level, and there is no such thing as a prevalence-corrected version of ICC.\footnote{There are empirical studies showing that ICC does depend on (probabilistic) prevalence \citep{gulli2005}. According to this study, ICC increases as prevalence increases towards $0.5$. This supports the explanation given here, because group 0 with higher scores is assigned a higher ICC level.}

\subsection{Experiment 2, Perturbation of All Features, $\sigma^2  = 5$} 
\label{exp_all_s5}

\begin{figure*}[!h] 
     \centering
     \begin{subfigure}{0.3\textwidth}
         \centering
         \includegraphics[width=\textwidth]{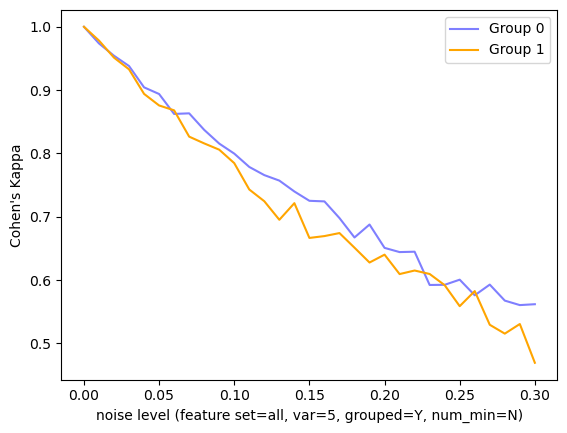}
         \caption{Cohen's Kappa}
         \label{fig:CK_s5}
     \end{subfigure}
     \hfill
     \begin{subfigure}{0.3\textwidth}
         \centering
         \includegraphics[width=\textwidth]{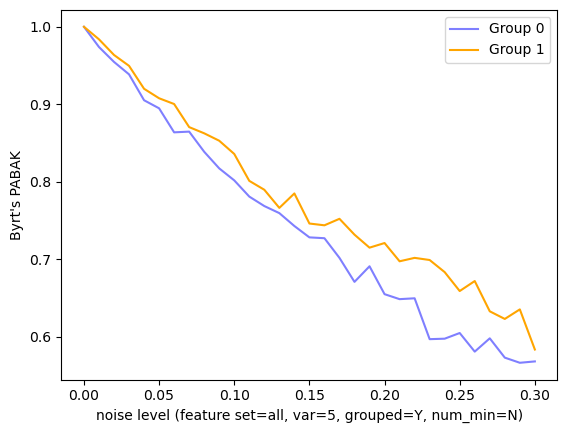}
         \caption{Byrt's PABAK}
         \label{fig:PABAK_s5}
     \end{subfigure}
          \hfill
     \begin{subfigure}{0.3\textwidth}
         \centering
         \includegraphics[width=\textwidth]{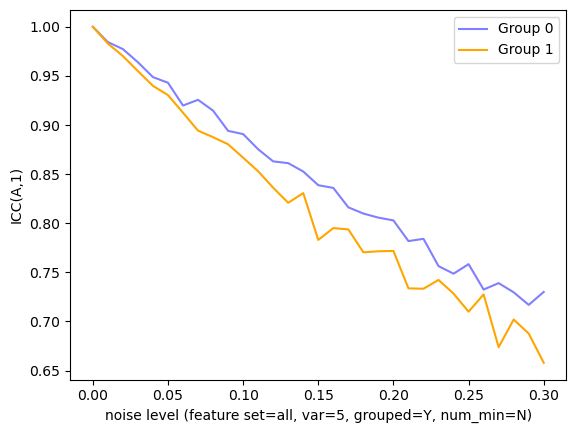}
         \caption{ICC(A,1)}
         \label{fig:ICCA_s5}
     \end{subfigure}
        \caption{Experiment 2, $\sigma^2 = 5$.}
\end{figure*}

In this experiment, the same features are perturbed, but with a larger variance of $5$ for numerical variables. Again, CK (Fig. \ref{fig:CK_s5}) decreases, but to a lower overall level than for $\sigma^2 = 1$ (Fig. \ref{fig:CK_s1}). This is what we expect: larger errors lead to lower reliability. We also observe a (small) systematic group difference: reliability is (mostly) lower for group 1. PABAK reliability (Fig. \ref{fig:PABAK_s5}) also decreases to a lower level than for $\sigma^2 = 1$ (Fig. \ref{fig:PABAK_s1}). Reliability is lower for group 0 than for group 1. The PABAK-$\Delta$ has opposite sign of CK (Fig. \ref{fig:CK_s5}). Again, we can examine how the difference between CK and PABAK came about by decomposing the PABAK correction into a bias correction and a prevalence correction. The result is the same as in experiment 1: the prevalence correction for group $1$ is mostly responsible for the difference between CK and PABAK. For plots see appendix \ref{sec:bias_prev_corr_ex2_3}. ICC (Fig. \ref{fig:ICCA_s5}) shows the same qualitative behavior as in experiment 1 (Fig. \ref{fig:ICCA_s1}).

\subsection{Experiment 3, Perturbation of All Features, $\sigma^2  = 10$}
\label{exp_all_s10}

\begin{figure*}[!h] 
     \centering
     \begin{subfigure}{0.3\textwidth}
         \centering
         \includegraphics[width=\textwidth]{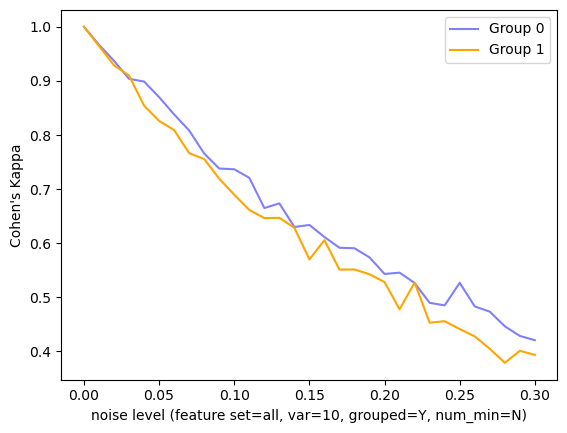}
         \caption{Cohen's Kappa}
         \label{fig:CK_s10}
     \end{subfigure}
     \hfill
     \begin{subfigure}{0.3\textwidth}
         \centering
         \includegraphics[width=\textwidth]{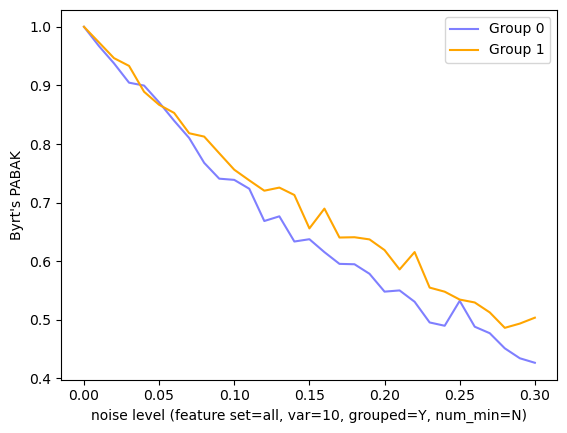}
         \caption{Byrt's PABAK}
         \label{fig:PABAK_s10}
     \end{subfigure}
          \hfill
     \begin{subfigure}{0.3\textwidth}
         \centering
         \includegraphics[width=\textwidth]{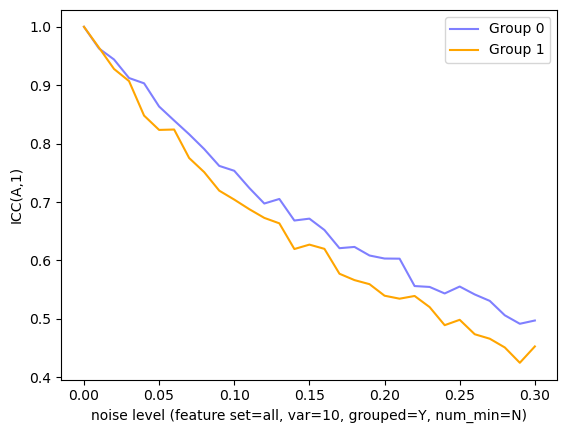}
         \caption{ICC(A,1)}
         \label{fig:ICCA_s10}
     \end{subfigure}
        \caption{Experiment 3, $\sigma^2 = 10$.}
\end{figure*}

Again, all features are perturbed; the variance for numerical features is increased to $10$. We observe the same general tendency in Fig. \ref{fig:CK_s10} as in experiments 1 (Fig. \ref{fig:CK_s1}) and 2 (Fig. \ref{fig:CK_s5}). CK descends to a low overall level. Reliability is systematically lower for group 1. Qualitatively, PABAK (Fig. \ref{fig:PABAK_s10}) shows a similar picture as in experiments 1 (Fig. \ref{fig:PABAK_s1}) and 2 (Fig. \ref{fig:PABAK_s5}). Overall levels are lower, and reliability of group 0 is below group 1. Again, we can examine how the difference between CK and PABAK came about. The reason is the same as in the first two experiments; for plots see appendix \ref{sec:bias_prev_corr_ex2_3}. ICC shows the same qualitative behavior as in experiments 1 and 2, with lower overall levels; group 1 has systematically lower reliability. Note that all three statistics show some sign of non-linearity as $\sigma^2$ grows; this is likely due to the numerical noise (normal distribution).

In sum, the different reliability statistics yield contradictory results as to which group has lower reliability due to the injection of uniform noise. On the one hand, Cohen's Kappa and ICC find a lover level of reliability for group $1$, putting this group at a disadvantage. Results become more pronounced as the error size ($\sigma^2$) grows. PABAK, on the other hand, finds lower reliability levels for group $0$. The differences between CK and PABAK can be explained by the prevalence correction in PABAK: Prevalence correction yields higher reliability levels for group $1$.

\subsection{Further Experiments}
\label{sec:further_experiments}

\paragraph{Perturbation of All Features with Minima for Numerical Features}

Fixed minima for the numerical features were tested in three experiments. If the perturbed value was below a minimum for a certain feature, it was reset to that minimum. The value of the feature ``age'' was set to a minimum of 18; the value of the ``counts'' features was set to a minimum of 0. The results are in appendix \ref{sec:all_features_num_minima}. The comparison with the results of the first three experiments shows a qualitative, with some quantitative differences. The overall level of reliability is slightly higher in the experiments with minima. Also, the PABAK gap closes as $\sigma^2$ is increased. This suggests that qualitatively, the results from the first three experiments do not depend on the specifics of the noise scheme.

\paragraph{Perturbation of Numerical Features}

Only numerical features were perturbed in three experiments; categorial features were not perturbed. The results are in appendix \ref{sec:perturb_num}. The results in these experiments are a bit noisier, in particular for $\sigma^2 = 1$. It can be observed that there is a systematic, qualitative difference between CK and PABAK here as well. For $\sigma^2 = 1$, the two groups have gaps with opposite sign. As $\sigma$ increases, the gap appears to become slightly more pronounced for CK, and smaller for PABAK. This shows that even if only the numerical features are perturbed, the qualitative finding is confirmed that there are reliability gaps between groups, and that the sign of the gap depends on the IRR statistic.

\paragraph{Perturbation of Categorical Features}

Only categorial features were perturbed in one experiment. The results can be found in appendix \ref{sec:perturb_cat}. Here it is observed that there appears to be no systematic CK gap between groups. The gap for PABAK is consistent with the first three experiments, but even more pronounced. This suggests that the PABAK gap we observe in the first three experiments may be driven by the categorial features to some extent, but not exclusively, as observed in the experiments with only numerical features.

\section{Discussion}
\label{sec:discussion}

\paragraph{Summary} 

The most important result is that there are reliability gaps between groups according to common IRR statistics (Cohen's Kappa, Byrt's PABAK, ICC(A, 1)), putting one or the other of the two groups (White people and non-White people) at a disadvantage. Qualitatively, we find that ICC(A,1), which measures the reliability of predicted scores, almost universally assigns a lower reliability to group 1. Cohen's Kappa, which measures the reliability of binary predictions, mostly finds the same result, with less consistency than ICC. Byrt's PABAK, on the other hand, which also measures the reliability of binary predictions, assigns lower reliability to group 0. We can account for these differences in the case of binary predictions (Cohen's Kappa vs. Byrt's PABAK): The bulk of the difference is due to the prevalence correction, measured by the prevalence index, which accounts for an underestimation of reliability for group 1 by Cohen's Kappa. The prevalence index mirrors differences in the distributions of $(Y, A)$, group-specific ground truth, because the predictor $f(X) = \hat{Y}$ is an approximation of the data $(X, A, Y)$, and the distribution of group-specific predictions $(\hat{Y}, A)$ reproduces the data to some extent.

\paragraph{Normative Upshot} 

A lack of reliability can be morally undesirable in its own right, because reliability can be interpreted as individual fairness, a notion requiring that similar people should get similar predictions; see \citet{raez2023a} for the full argument. A consequence of this interpretation is that if a group has lower reliability, the representations of its members are less adequate for predictive purposes. 

It could be asked whether a reliability gap constitutes discrimination.\footnote{\citet{lippe2013} is a standard philosophical account of discrimination.} It is possible that a rater intentionally treats members of different groups differently due to their group membership, which constitutes direct discrimination. However, it can also happen that differences in reliability are due to the fact that different groups have different feature distributions, such that mistakes have different consequences for different groups. These differences are due to historical bias in the data, not to biased ratings. Of course, such differences may need to be mitigated, but they do not constitute direct discrimination by the raters. 

Reliability gaps are relevant not only in the context of RAIs, but also in the context of medicine and other disciplines. The normative significance of reliability may be context dependent. In the context of RAIs, base rates are due to historical inequalities and thus in need of mitigation. In other contexts, base rates may have different sources, which means that different or no mitigation is adequate.

\paragraph{Methodological Ramifications} 

In empirical research, reliability is usually measured by Cohen's Kappa, and by versions of ICC. The above findings suggest that this may be problematic in recidivism risk assessment, because Cohen's Kappa strongly depends on prevalences, which, in turn, mirror historical inequalities between groups. Specifically, CK underestimates reliability for groups with lower prevalence -- in the present case, the privileged group 1. It is therefore suggested that reliability research should also report base rates (in the form of prevalence and bias index), and measure PABAK besides CK to gauge the extent of prevalence and bias dependence of CK. We also saw that ICC almost universally found group 1 to have lower reliability. This finding is mostly consistent with reliability as measured by Cohen's Kappa. This, of course, suggests that  ICC depends on prevalence with respect to predicted scores. It was noted above that empirical results suggest a prevalence dependence of ICC. However, the author is not aware of prevalence corrected versions of ICC. Understanding this issue better from a theoretical (statistical) point of view is important, in particular because reliability of scores is almost exclusively evaluated with ICC.

\paragraph{Simulation vs. Empirical Approach}

The above experiments show that at the proof-of-concept level, using simulated IRR data is a promising approach that can complement experimental measurements of IRR. Simulated data can be generated and investigated at no cost to judicial agencies. Also, the simulation approach used here takes away the emphasis from the (un-)reliability of human raters and puts it on the reliability of instruments, and in particular on the features used by RAIs, because synthetic ``raters'' have transparent properties (noise scheme). The simulation approach also provides a richer, more qualitative picture of possible reliability gaps by not providing point measurements, but the behavior of IRR measures on an entire range. 

This should not be read as a plea against real inter-rater data. Rather, inter-rater data can be used in combination with simulations. For example, if inter-rater data for a representative but small sample is available, one can use this limited dataset to create a simulation study of a larger dataset for the same instrument, by creating synthetic raters (noise) based on the error profile of the raters from the empirical inter-rater sample. The real inter-rater data would provide empirical information about the distribution of errors for the different features; the simulation study would reproduce these empirical distributions, but could vary the noise level to some extent, as in the experiments performed above. In this way, the strengths of the empirical and the simulation approach can be combined. An approach using empirical information about noise distributions would be preferable to the approach taken above, where the noise scheme is somewhat \emph{ad hoc}. The use of normal noise in the case of numerical features is not particularly realistic, and results should be taken with a grain of salt. This approach can only provide a qualitative idea of how errors in numerical features affect inter-rater reliability. We have also seen that there appears to be a qualitatively different behavior of reliability with respect to some of the IRR statistics. It cannot be excluded that this is due to the noise scheme used here.

\paragraph{Mitigation}

The above experiments and discussion focus on the identification of possible reliability gaps between socially salient groups. Little has been said about mitigation. One lesson is that, generally speaking, the very same level of noise can impact different groups of people differently. This alone provides a further, fairness-driven motivation to better understand and improve reliability of RAIs, besides the usual argument that reliability is necessary for predictive validity. One plausible way to increase the general level of reliability is via sparse models. This approach has been advocated and tested to improve predictive validity. For example, \citet{dress2018} have argued that a model using two features has the same accuracy as COMPAS. It is very likely that sparsity would also improve reliability, because fewer features also means fewer opportunities for mistakes. On the other hand, it is still important to carefully test this approach, because if fewer features are used, these at least prima facie also become more important, and so do mistakes in these features. Generally speaking, different model architectures should be investigated with respect to their reliability, both on real and simulation data.

\section{Conclusion}
\label{sec:conclusion}

In simulation studies based on the COMPAS dataset, it was shown that there are inter-rater reliability gaps between socially salient groups. According to common statistical measures (Cohen's Kappa, ICC(A,1)), White people have a lower level of reliability for the same level of noise. However, according to another statistical measure (Byrt's PABAK), which controls for different prevalences of the groups involved, non-White people have a lower level of reliability. One important recommendation for empirical research is that prevalence (and bias) should be reported when measuring IRR with Cohen's Kappa and ICC. Important open issues include the integration of empirical error distributions in the noise scheme, the exploration of different possibilities to mitigate large reliability losses, and theoretical work on the prevalence dependence of reliability statistics (ICC).

\onecolumn

\appendix

\section*{Appendix}

\section{Bias and Prevalence Correction, Experiments 2 and 3}
\label{sec:bias_prev_corr_ex2_3}

\begin{figure*}[h] 
     \centering
     \begin{subfigure}{0.4\textwidth}
         \centering
         \includegraphics[width=\textwidth]{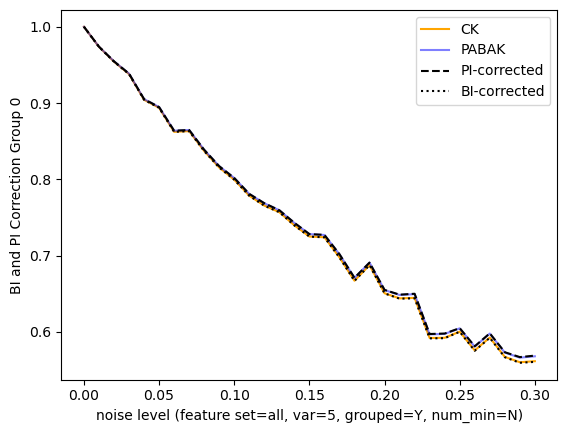}
         \caption{Group 0}
         \label{fig:corr_g0_s5}
     \end{subfigure}
     \hfill
     \begin{subfigure}{0.4\textwidth}
         \centering
         \includegraphics[width=\textwidth]{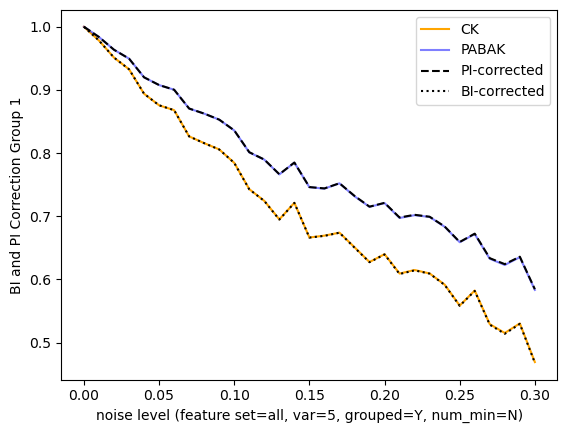}
         \caption{Group 1}
         \label{fig:corr_g1_s5}
     \end{subfigure}
        \caption{PI and BI correction experiment 2, $\sigma^2 = 5$.}
\end{figure*}

\begin{figure*}[h] 
     \centering
     \begin{subfigure}{0.4\textwidth}
         \centering
         \includegraphics[width=\textwidth]{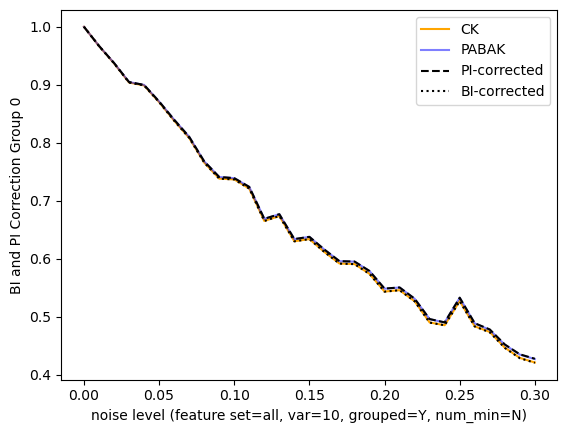}
         \caption{Group 0}
         \label{fig:corr_g0_s10}
     \end{subfigure}
     \hfill
     \begin{subfigure}{0.4\textwidth}
         \centering
         \includegraphics[width=\textwidth]{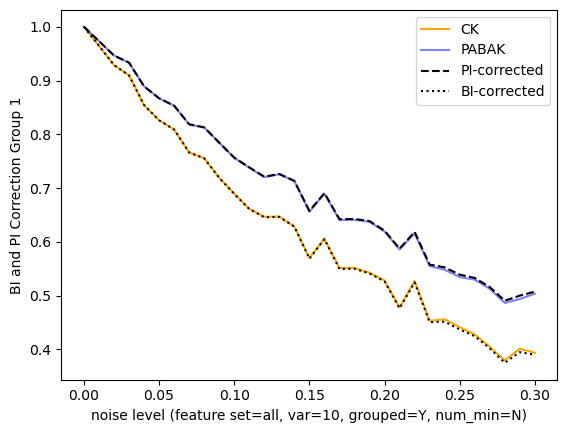}
         \caption{Group 1}
         \label{fig:corr_g1_s10}
     \end{subfigure}
        \caption{PI and BI correction experiment 3, $\sigma^2 = 10$.}
\end{figure*}

\newpage

\section{Perturbation of All Features, Minima for Numerical Features}
\label{sec:all_features_num_minima}

\begin{figure*}[!h] 
     \centering
     \begin{subfigure}{0.3\textwidth}
         \centering
         \includegraphics[width=\textwidth]{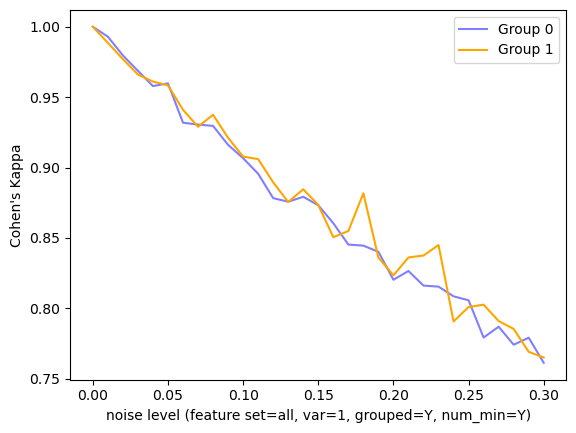}
         \caption{Cohen's Kappa, $\sigma^2 = 1$.}
         \label{fig:CK_s1_min}
     \end{subfigure}
     \hfill
     \begin{subfigure}{0.3\textwidth}
         \centering
         \includegraphics[width=\textwidth]{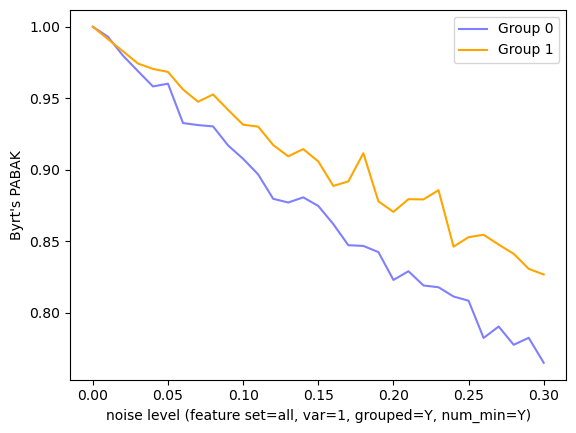}
         \caption{Byrt's PABAK, $\sigma^2 = 1$.}
         \label{fig:PABAK_s1_min}
     \end{subfigure}
     \hfill
     \begin{subfigure}{0.3\textwidth}
         \centering
         \includegraphics[width=\textwidth]{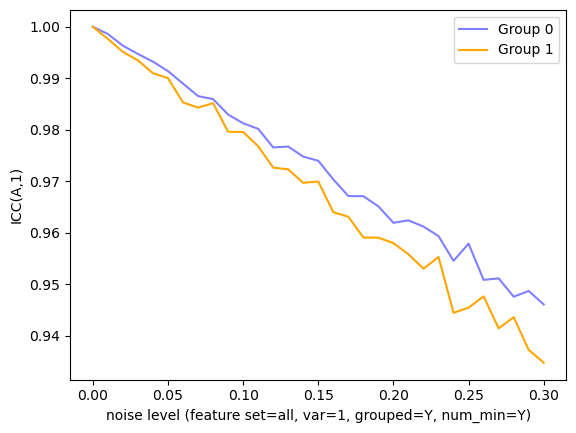}
         \caption{ICC(A,1), $\sigma^2 = 1$.}
         \label{fig:ICCA_s1_min}
     \end{subfigure}
     
     \bigskip
     \begin{subfigure}{0.3\textwidth}
         \centering
         \includegraphics[width=\textwidth]{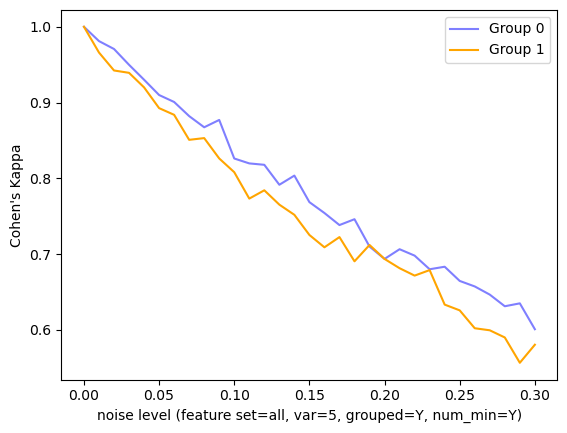}
         \caption{Cohen's Kappa, $\sigma^2 = 5$.}
         \label{fig:CK_s5_min}
     \end{subfigure}
     \hfill
     \begin{subfigure}{0.3\textwidth}
         \centering
         \includegraphics[width=\textwidth]{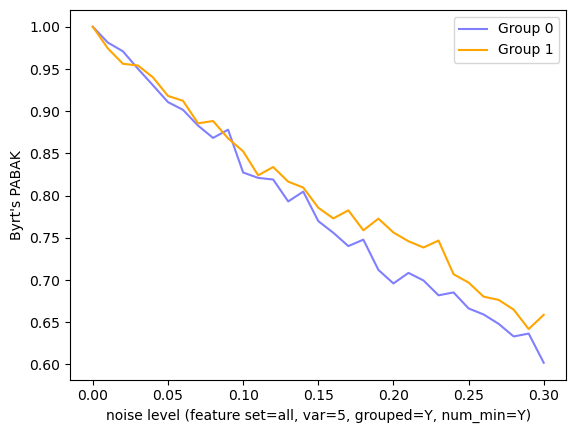}
         \caption{Byrt's PABAK, $\sigma^2 = 5$.}
         \label{fig:PABAK_s5_min}
     \end{subfigure}
     \hfill
     \begin{subfigure}{0.3\textwidth}
         \centering
         \includegraphics[width=\textwidth]{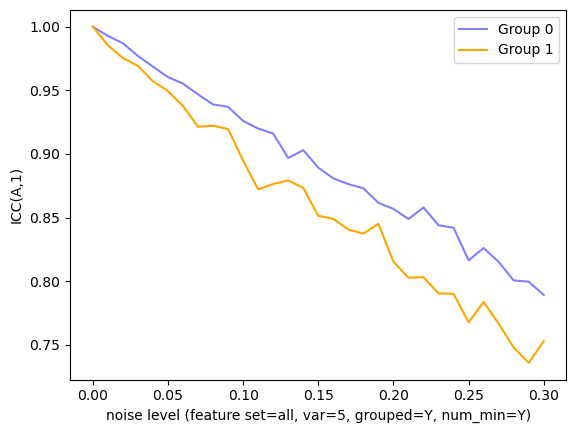}
         \caption{ICC(A,1), $\sigma^2 = 5$.}
         \label{fig:ICCA_s5_min}
     \end{subfigure}
     \bigskip     

     \begin{subfigure}{0.3\textwidth}
         \centering
         \includegraphics[width=\textwidth]{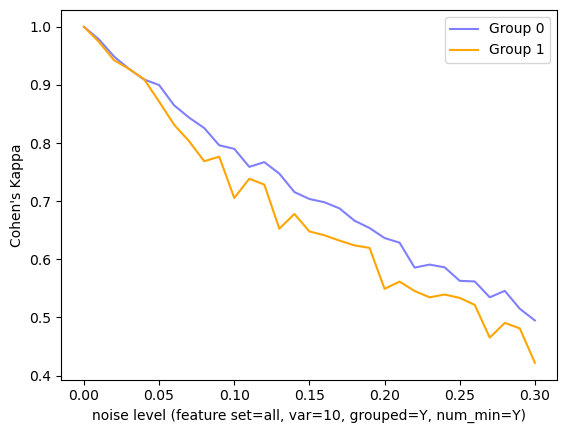}
         \caption{Cohen's Kappa, $\sigma^2 = 10$.}
         \label{fig:CK_s10_min}
     \end{subfigure}
     \hfill
     \begin{subfigure}{0.3\textwidth}
         \centering
         \includegraphics[width=\textwidth]{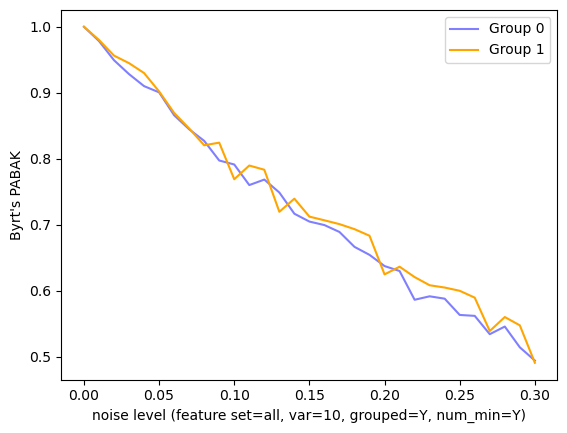}
         \caption{Byrt's PABAK, $\sigma^2 = 10$.}
         \label{fig:PABAK_s10_min}
     \end{subfigure}
          \hfill
     \begin{subfigure}{0.3\textwidth}
         \centering
         \includegraphics[width=\textwidth]{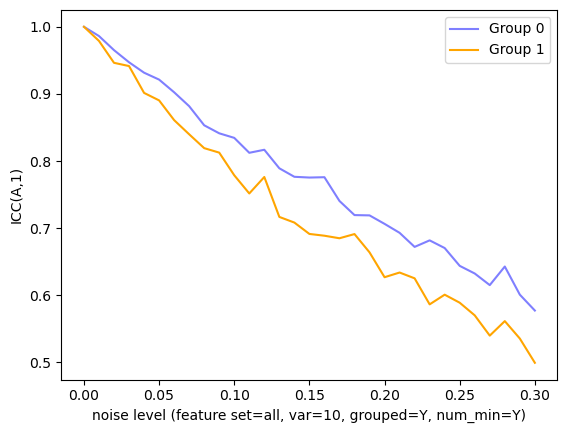}
         \caption{ICC(A,1), $\sigma^2 = 10$.}
         \label{fig:ICCA_s10_min}
     \end{subfigure}
        \caption{Three experiments with perturbation of all features and minima for numerical features. Each row corresponds to one experiment. Row 1: $\sigma^2 = 1$; row 2: $\sigma^2 = 5$; row 3: $\sigma^2 = 10$.}
\end{figure*}

\newpage

\section{Perturbation of Numerical Features}
\label{sec:perturb_num}

\begin{figure*}[!h] 
     \centering
     \begin{subfigure}{0.3\textwidth}
         \centering
         \includegraphics[width=\textwidth]{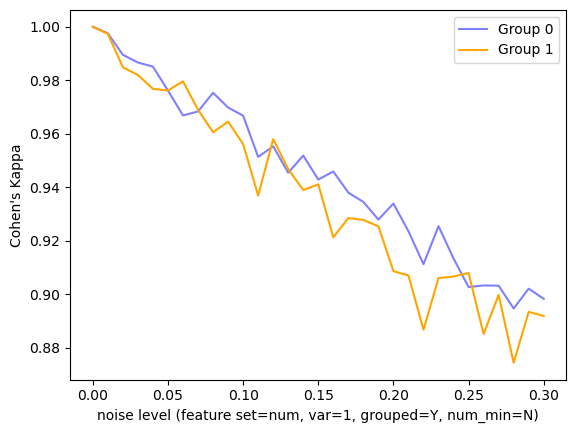}
         \caption{Cohen's Kappa, $\sigma^2 = 1$.}
         \label{fig:CK_s1_num}
     \end{subfigure}
     \hfill
     \begin{subfigure}{0.3\textwidth}
         \centering
         \includegraphics[width=\textwidth]{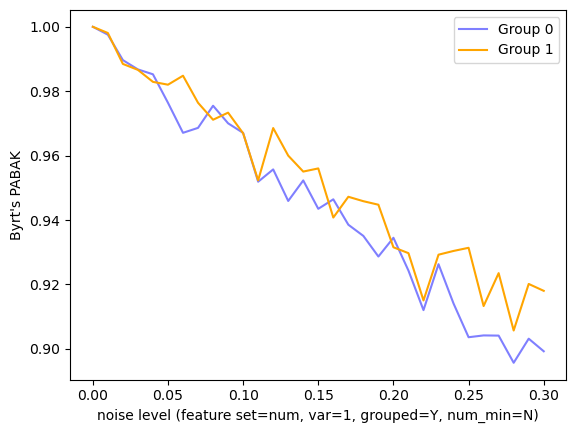}
         \caption{Byrt's PABAK, $\sigma^2 = 1$.}
         \label{fig:PABAK_s1_num}
     \end{subfigure}
          \hfill
     \begin{subfigure}{0.3\textwidth}
         \centering
         \includegraphics[width=\textwidth]{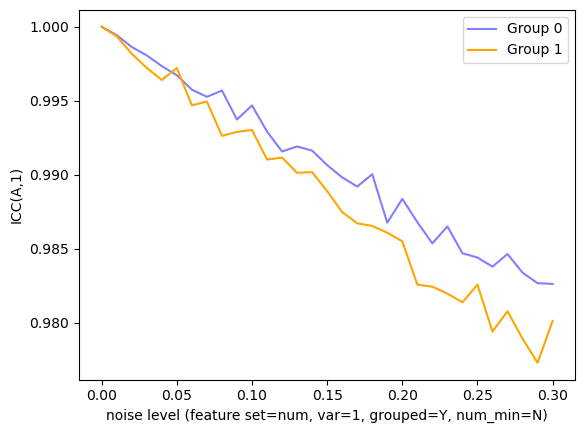}
         \caption{ICC(A,1), $\sigma^2 = 1$.}
         \label{fig:ICCA_s1_num}
     \end{subfigure}
     \bigskip

     \begin{subfigure}{0.3\textwidth}
         \centering
         \includegraphics[width=\textwidth]{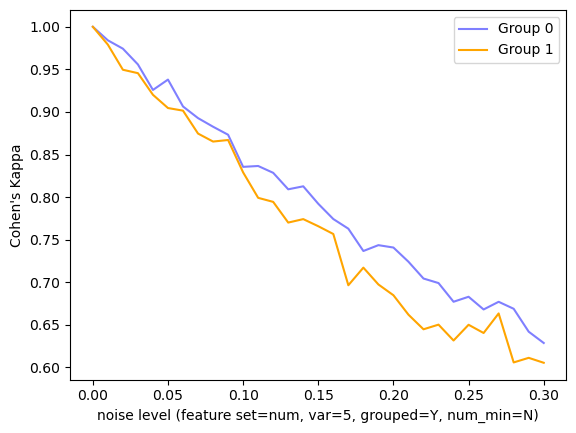}
         \caption{Cohen's Kappa, $\sigma^2 = 5$.}
         \label{fig:CK_s5_num}
     \end{subfigure}
     \hfill
     \begin{subfigure}{0.3\textwidth}
         \centering
         \includegraphics[width=\textwidth]{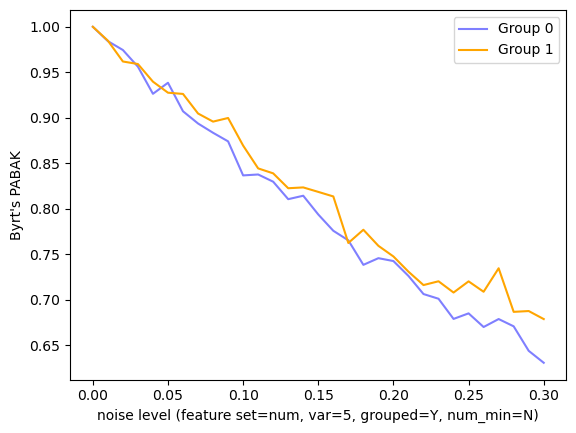}
         \caption{Byrt's PABAK, $\sigma^2 = 5$.}
         \label{fig:PABAK_s5_num}
     \end{subfigure}
          \hfill
     \begin{subfigure}{0.3\textwidth}
         \centering
         \includegraphics[width=\textwidth]{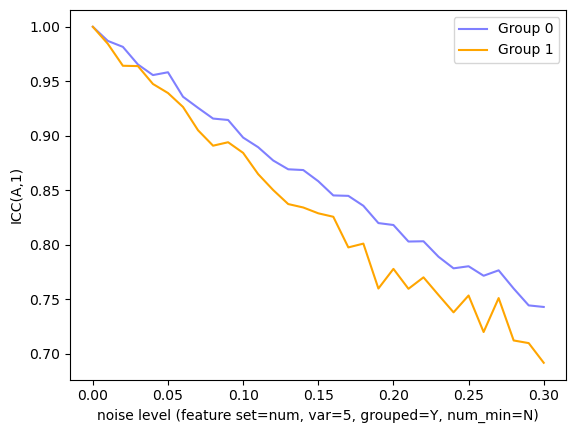}
         \caption{ICC(A,1), $\sigma^2 = 5$.}
         \label{fig:ICCA_s5_num}
     \end{subfigure}
     \bigskip
    
     \centering
     \begin{subfigure}{0.3\textwidth}
         \centering
         \includegraphics[width=\textwidth]{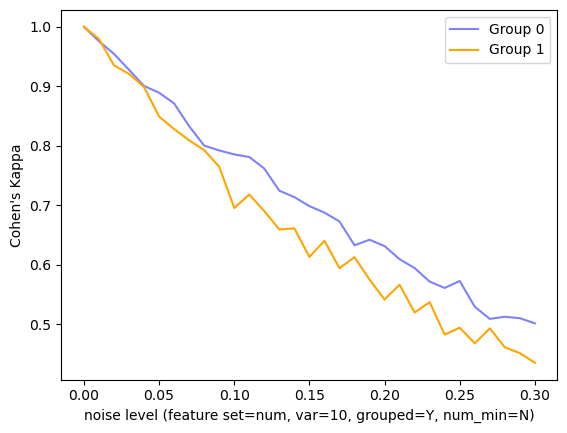}
         \caption{Cohen's Kappa, $\sigma^2 = 10$.}
         \label{fig:CK_s10_num}
     \end{subfigure}
     \hfill
     \begin{subfigure}{0.3\textwidth}
         \centering
         \includegraphics[width=\textwidth]{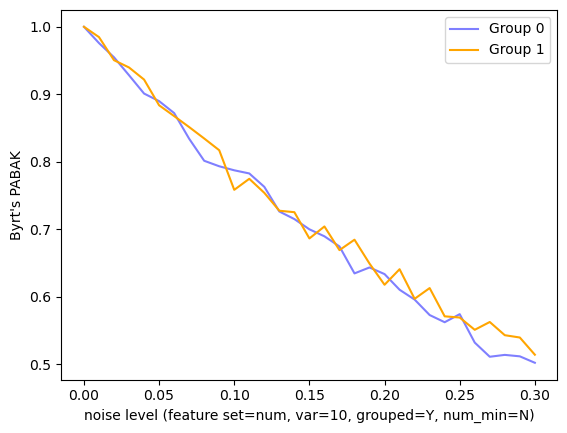}
         \caption{Byrt's PABAK, $\sigma^2 = 10$.}
         \label{fig:PABAK_s10_num}
     \end{subfigure}
          \hfill
     \begin{subfigure}{0.3\textwidth}
         \centering
         \includegraphics[width=\textwidth]{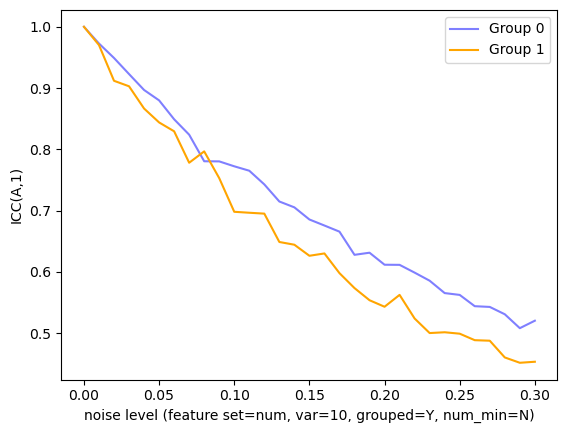}
         \caption{ICC(A,1), $\sigma^2 = 10$.}
         \label{fig:ICCA_s10_num}
     \end{subfigure}
        \caption{Three experiments with perturbation of numerical features only. Each row corresponds to one experiment. Row 1: $\sigma^2 = 1$;  row 2: $\sigma^2 = 5$; row 3: $\sigma^2 = 10$.}
\end{figure*}

\newpage

\section{Perturbation of Categorical Features}
\label{sec:perturb_cat}

\begin{figure*}[!h] 
     \centering
     \begin{subfigure}{0.3\textwidth}
         \centering
         \includegraphics[width=\textwidth]{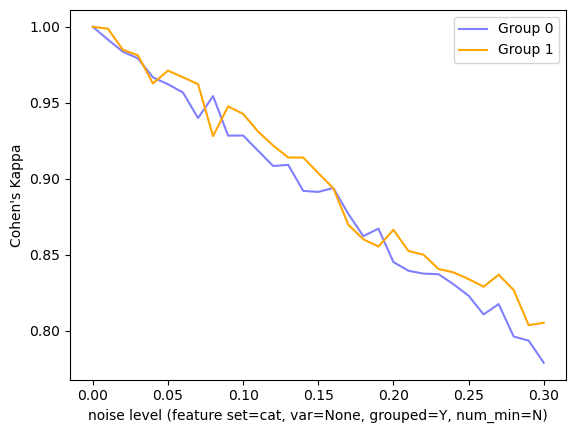}
         \caption{Cohen's Kappa}
         \label{fig:CK_cat}
     \end{subfigure}
     \hfill
     \begin{subfigure}{0.3\textwidth}
         \centering
         \includegraphics[width=\textwidth]{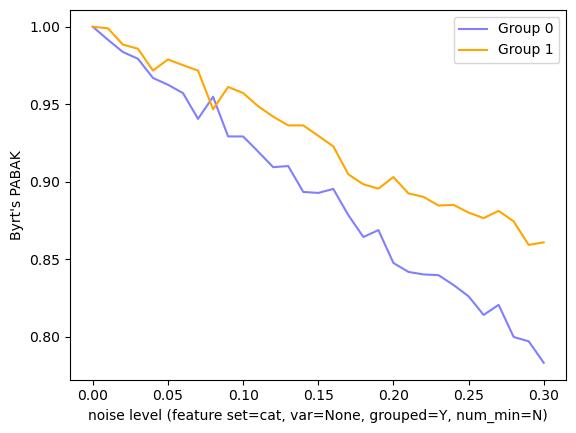}
         \caption{Byrt's PABAK}
         \label{fig:PABAK_cat}
     \end{subfigure}
          \hfill
     \begin{subfigure}{0.3\textwidth}
         \centering
         \includegraphics[width=\textwidth]{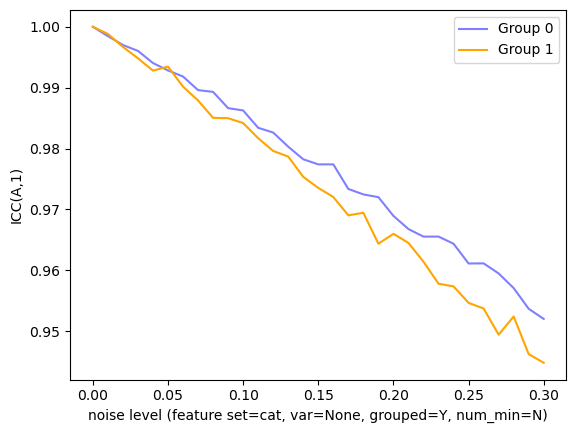}
         \caption{ICC(A,1)}
         \label{fig:ICCA_cat}
     \end{subfigure}
        \caption{Experiment with perturbation of categorical features only.}
\end{figure*}

\end{document}